\documentclass[pra,twocolumn]{revtex4}

\usepackage{amsmath}                          
\usepackage{amsfonts}                         
\usepackage{amssymb}                          
\usepackage{amscd}                            
\usepackage{amsthm}                           

\usepackage{tikz}
\usepackage{tkz-graph}
\usepackage{tkz-berge}
\usepackage[utf8]{inputenc}
\usepackage{dsfont}

\usepackage{dsfont}                           

\usepackage{graphicx}                         
\usepackage{epstopdf}                         
\usepackage{subfigure}			      

\usepackage[colorlinks=true,
            urlcolor=blue,
            citecolor=blue,
            linkcolor=black,
            pdfstartview=FitH,
            pdfpagemode=None]{hyperref}

\begin{document}

\title{Homological Stabilizer Codes}

\author{Jonas T. \surname{Anderson}}
\email[]{jander10@unm.edu}
\affiliation{Center for Quantum Information and Control,
             University of New Mexico,
             Albuquerque, NM, 87131, USA}


\begin{abstract}
In this paper we define homological stabilizer codes which encompass codes such as Kitaev's toric code and the topological color codes. These codes are defined solely by the graphs they reside on. This feature allows us to use properties of topological graph theory to determine the graphs which are suitable as homological stabilizer codes. We then show that all toric codes are equivalent to homological stabilizer codes on 4-valent graphs. We show that the topological color codes and toric codes correspond to two distinct classes of graphs. We define the notion of label set equivalencies and show that under a small set of constraints the only homological stabilizer codes without local logical operators are equivalent to Kitaev's toric code or to the topological color codes.
\end{abstract}

\maketitle

\section{Introduction}

Stabilizer codes \cite{Gottesman:1997a} offer protection from bounded-weight qubit errors using measurements of qubit Pauli operators. Surface codes such as the well-known toric code \cite{Kitaev:1997a, Dennis:2002a} can be realized with a two dimensional layout. More recently the color codes \cite{Bombin:2006b} were introduced. They have similar properties as the toric code with additional transversal gates. The locality of stabilizer generators and high error threshold \cite{Dennis:2002a, Raussendorf:2007a, Wang:2003a, Katzgraber:2009a, Andrist:2010a, Anderson:2011a} make these codes some of the most promising for use as a quantum memory. 

These codes can be used as a standard quantum memory or we can make ``holes'' in the surface and braid these holes around each other to implement robust topological logic gates  \cite{Raussendorf:2006a, Raussendorf:2007a, Raussendorf:2007b, Bombin:2007e}. The homological codes are the only known codes with this property. 

Kitaev's toric code (KTC), the topological color code (TCC), and the Levin-Wen plaquette model (LWPM) \cite{LW:2003a} are all different homological codes. All of these codes have stabilizer generators that correspond to faces on a graph. In this paper we start by identifying stabilizer generators with faces on a planar graph embedded in some surface. Then using techniques from graph theory we find the lattices upon which we can construct these codes and their generalizations. We call these codes homological stabilizer codes (HSC) after a similar class of codes discovered by Bombin et al. \cite{Bombin:2007c}. All known surface stabilizer codes are shown to be special cases of this general structure. 

The known homological stabilizer codes \cite{Kitaev:2005a, Levin:2006a, WenBook:2007a} provide exactly-solvable models with exhibit topological order. These models allow us to probe topological order and, while simple, have many interesting properties and are currently an active area of study. Finding new models with topological order would increase our understanding of how topological order arises. 

First, we will introduce the theory of graphs in the abstract. Homological properties of these graphs will be emphasized. Then, stabilizer codes will be introduced. KTC and the TCCs are then presented in a graph-theoretic context. We will then give a set of conditions that a planar graph must have to admit a HSC. Finally, we prove that all HSC that have no local logical operators are equivalent to the TCCs or to KTC. In light of the recent result by Bombin et al. \cite{Bombin:2011a}, we can say that all 2D HSCs are equivalent to one or two copies of KTC. The topological phases of 2D stabilizer codes with no local logical operators and translationally invariant stabilizers were studied by Yoshida\cite{Yoshida:2011a}. The topological phases of these systems were classified by their logical operators. 

\section{Graph Theory}

In this section we will introduce the basics of graph theory \cite{Diestel:1997, BondyMurty:2008} with emphasis on surface embeddings. 

A \emph{graph} G is a set of vertices ($V(G)$) and edges ($E(G)$). More specifically,
$G$ is defined as an ordered pair $(V(G),E(G))$. Elements of $E(G)$ can be defined by the 
two vertices they connect for example: $\{v_a, v_b\} \in E(G)$. We will only be discussing undirected
graphs where $\{v_{a}, v_{b}\} = \{v_{b}, v_{a}\}$. Simple graphs are undirected graphs without self-loops $\{v_{a}, v_{a}\}$ or multi-edges.

A \emph{path} on a graph is a set of ordered vertices (or edges) such that each vertex (edge) is connected (shares a vertex) with the next. 
A \emph{cycle} on a graph is a path that returns to its starting vertex without crossing itself. A \emph{facial cycle} is a cycle 
that corresponds to a face. 

A graph is connected if there is a path between any two vertices.
A graph is said to be $k$-connected if removal of $k-1$ vertices and their corresponding edges
leaves the graph connected. In this paper we will mainly be interested in properties of $(k\ge2)$-connected graphs.

Planar graphs are graphs that can be embedded in the $2D$ Euclidean plane or the sphere. A graph is embeddable in a surface if the graph 
can be drawn on the surface such that the edges of the graph touch only at vertices.

A {\em subdivision} of an edge involves adding one or more vertices to an edge. A subdivision of a graph is a graph with a set of subdivided edges. 

A {\em subgraph} of a graph $G$ is a graph that can be obtained by repeated edge and vertex deletions on $G$. 

A necessary and sufficient condition for a graph to be planar is given by Kuratowski's Theorem: \textit{A graph is 
planar iff it does not contain a subgraph that has $K_{5}$ or $K_{3,3}$ as subdivisions.} See Fig.~\ref{fig:k33},\ref{fig:k5}

\begin{figure}
\begin{tikzpicture}
  \SetVertexNormal[Shape =  circle,
                   FillColor  = black,
                   LineWidth  = 4pt]
  \SetUpEdge[lw         = 2.5pt,
             color      = black,
             labelcolor = white,
             labeltext  = red,
             labelstyle = {sloped,draw,text=blue}]
 \Vertex[x=0, y=0]{G}
 \Vertex[x=0, y=2]{A} 
 \Vertex[x=0, y=4]{P}
 \Vertex[x=3, y=0]{C}
 \Vertex[x=3, y=2]{Q}
 \Vertex[x=3, y=4]{E}
 \Edges(G,C) \Edges(A,Q) \Edges(P,E)\Edges(G,Q)\Edges(G,E)\Edges(A,C)\Edges(A,E)\Edges(P,Q)\Edges(P,C) 
\end{tikzpicture}
\caption{\label{fig:k33}$K_{3,3}$}
\end{figure}

\begin{figure}
\begin{tikzpicture}
\GraphInit[vstyle=Shade]
  \SetVertexNormal[Shape      = circle,
                   FillColor  = black,
                   LineWidth  = 4pt]
  \SetUpEdge[lw         = 2.5pt,
             color      = black,
             labelcolor = white,
             labeltext  = red,
             labelstyle = {sloped,draw,text=blue}]
\grComplete[Math,RA=2]{5}
\end{tikzpicture}
\caption{\label{fig:k5}$K_{5}$}
\end{figure}

A graph is planar graphs iff it obeys the Euler formula:
\begin{equation}
 |F(G)|-|E(G)|+|V(G)|=2.
\end{equation}
Where $|F(G)|$ is the number of faces, $|E(G)|$ is the number of edges, and $|V(G)|$ is the number of vertices of the graph $G$. We will 
usually drop the $G$ when it is clear from context. 
When drawing planar graphs the outer boundary or unbounded face is also considered a face. The RHS of the above equation is called the Euler characteristic.
The Euler characteristic is $2$ for all planar graphs (including non-simple graphs). Since the surface of a sphere with one pole removed can be projected to the 2D plane through stereographic projection, the Euler characteristic of a planar graph and a graph embeddable in the sphere are equivalent. 

\begin{figure}[htb]
\center{\includegraphics[width=0.9\columnwidth]{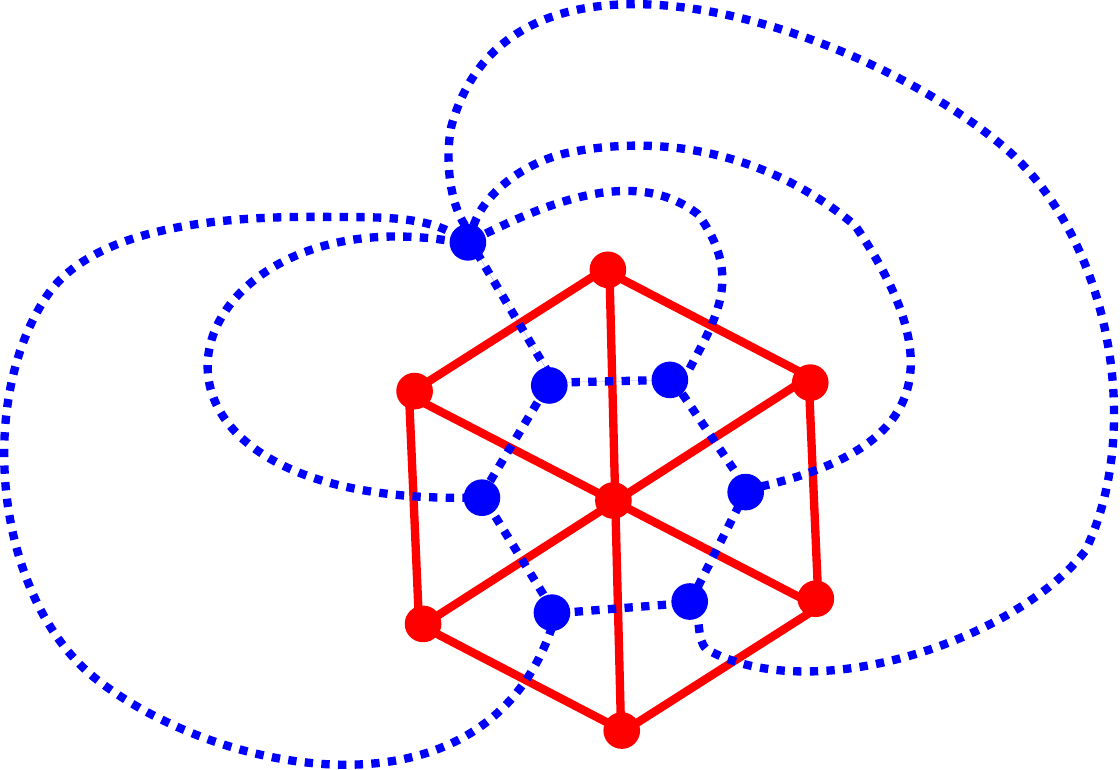}}
\caption{\label{fig:PrimalDual} The red is the primal graph and blue represents its dual graph. The dual graph of a dual graph is always the original graph.}
\end{figure}

The dual of $G$ denoted $G^{*}$ maps vertices from $G$ to faces in $G^*$ and faces in $G$ to vertices in $G^*$. The number of edges $|E|$ is unaffected by the dual transformation and hence the Euler characteristic is unchanged under the dual transformation see Fig.~\ref{fig:PrimalDual}. This proves that the dual of a planar graph is always planar. The dual of a dual graph applies the identity to a graph, therefore, the dual is its own inverse. 

\begin{figure}[htb]
\center{\includegraphics[width=0.6\columnwidth]{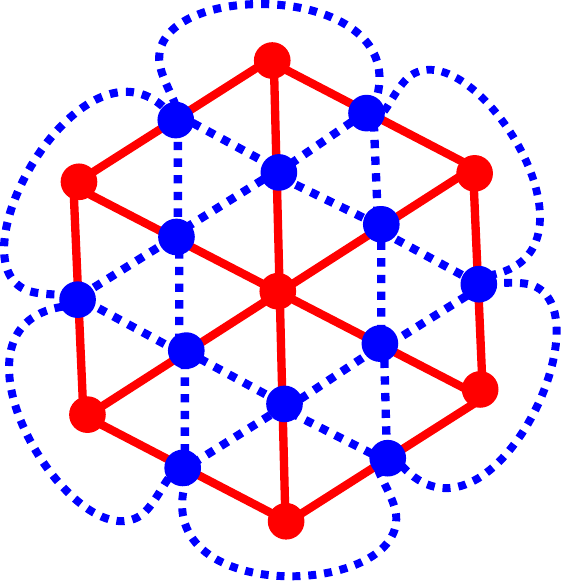}}
\caption{\label{fig:PrimalMedial} Again, the red graph is the primal graph. Now, blue represents its medial graph. The medial graph of a planar graph is always 4-valent.}
\end{figure}

The medial graph denoted $G^{M}$ maps edges in $G$ to vertices in $G^{M}$. The medial transformation puts a vertex at each edge in $G$, then connects vertices with an edge if the vertices are neighbors in the same face see Fig.~\ref{fig:PrimalMedial}. The faces and vertices of $G$ are mapped to faces in $G^{M}$. 2-connected planar graphs have edges which are part of exactly two faces. Each edge in these graphs will be mapped to a 4-valent vertex. The medial graph will be non-simple iff the original planar graph has one or two-valent vertices. The one-valent vertices 
become self-loops while the two-valent vertices become two-edges under the medial transformation. These non-simple graphs will still be 4-valent. Therefore, all planar graphs are mapped to 4-valent graphs under the medial transformation. Below is a proof that the Euler characteristic is unchanged under the medial transformation:

The number of edges in a graph is related to the number of vertices by 
\begin{equation}
2|E|=3|V_{3}|+4|V_{4}|+5|V_{5}|+...
\end{equation} 
Where $|V_{n}|$ is the number of $n$-valent vertices in $G$.  

From the above equation we can easily see that the average valency can be expressed as
\begin{equation}
avg val(G)=\frac{2|E|}{|V|}
\end{equation} 

Each vertex in a planar graph will be four-valent after a medial transformation.
Therefore,$|V(G_{M})|=V_{4}$, and $|E(G_{M})|=2|V(G_{M})|=2|E(G)|$. Therefore, the medial graph has $2|E(G)|$ edges.
After, the medial graph transformation we have: 
\begin{equation*}
\begin{split}
 |V(G)|-|E(G)|+|F(G)|  &=\\
 |E(G)|-2|E(G)|+(|V(G)|+|F(G)|) &=\\ 
 |V(G_{M})|-|E(G_{M})|+|F(G_{M})| &
\end{split}
\end{equation*}	
Hence, the Euler characteristic remains unchanged for planar graphs after a medial transformation. Therefore, the medial transformation
does not affect planarity.

\begin{center}
\begin{tabular}{|c|c|c|}
    \hline
    Primal & Dual & Medial\\
    \hline \hline
    Face (n-sided) &  Vertex (n-valent) & Face (n-sided) \\
    Edge & Edge & Vertex \\
    Vertex(m-valent) & Face (m-sided) & Face (m-sided) \\ 
    \hline
\end{tabular}
\end{center}	
Planar 4-valent graphs are always two-face-colorable. 

The set of all cycles in a graph is known as the cycle space denoted $Z(G)$. When $G$ is a 2-connected, planar graph,
the set of all simple or facial cycles of $G$ generates the cycle space $Z(G)$. As can easily be shown any one facial cycle can be removed and the 
resulting set forms a generating basis for $Z(G)$. Without loss of generality we can choose this cycle to be the outer unbounded face of the graph.

The set of all cuts in a graph generates the cut space denoted $C^{*}(G)$. A cut is a set of edges $E(G)$ in a connected graph $G$ that form a cycle in the dual graph. A cut vertex is the set of edges incident on a vertex. This creates a partition that isolates the cut vertex from the rest of the graph. These are the cut space analogue of facial cycles. For 2-connected, planar graphs the set of edges corresponding to each cut vertex generates the cut space. Again we can see that there will be one cut that can be expressed as product of the other cuts. 

\begin{figure}[htb]
\center{\includegraphics[width=0.6\columnwidth]{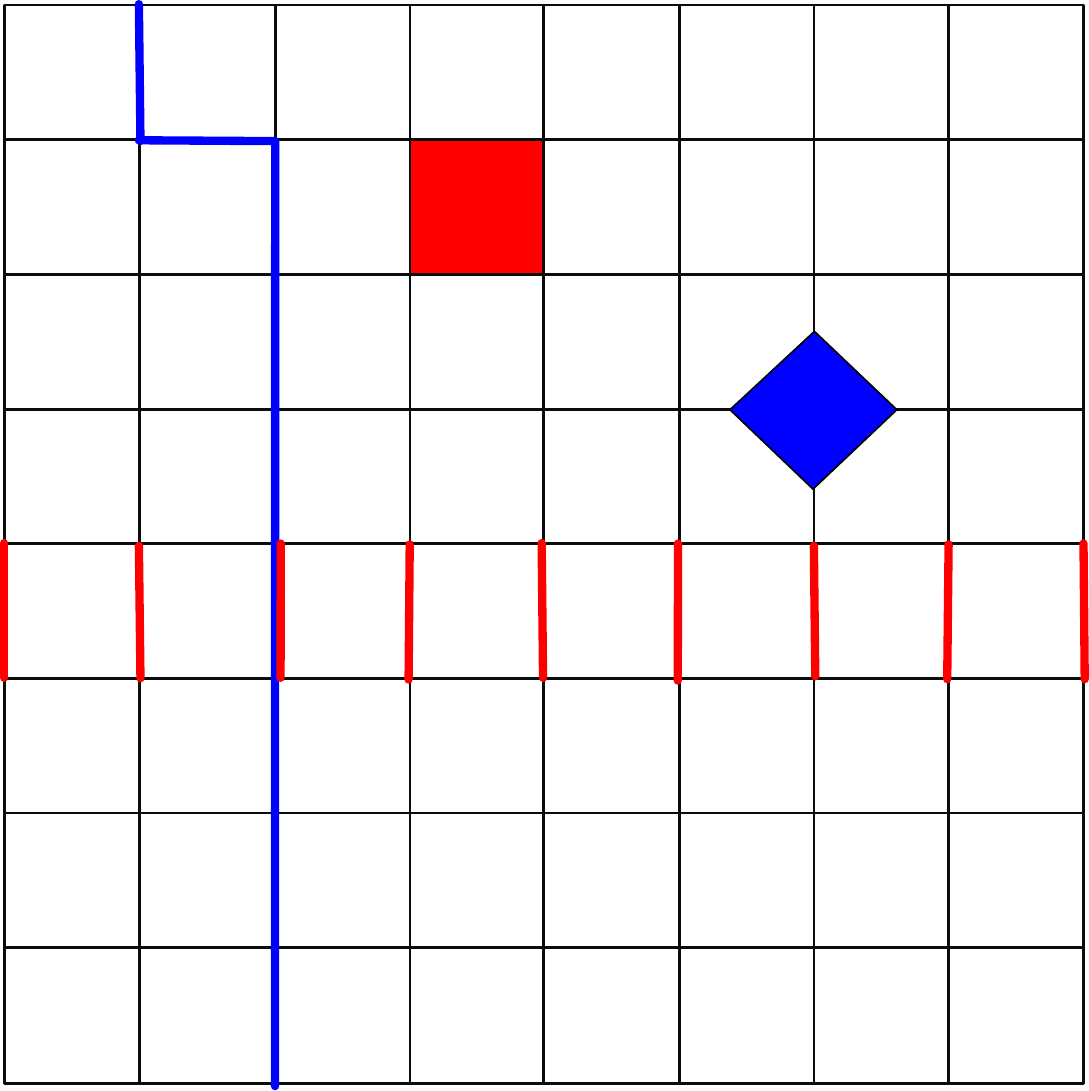}}
\caption{(a) a simple or facial cycle. (b) a simple or vertex cut. Notice that a simple cut is a simple cycle on the dual graph.} 
\end{figure}

Another equally good way to classify planar graphs is by their cycle and cut spaces. A graph $G$ is planar iff the cycle space of $G$ is equal to the cut space of the dual graph $G^{*}$. Or 

\begin{equation}
Z(G)=C^{*}(G^{*}).
\end{equation} 
  
We can now express the effect of dual and medial graph transformation in terms of the effects on the cycle and cut spaces. The medial graph transformation takes faces and vertices in the primal graph and takes them to faces in the medial graph. Or in terms of cuts and cycles the simple cuts as well as the simple cycles in $G$ will both be transformed to simple cycles in the medial graph $G^{M}$. The dual and primal planar graphs have been
mapped to a single planar graph. In addition, the resulting planar graph is a 4-valent graph which is always 2-face-colorable. The simple cycles in $G$ will be mapped to one color of faces in $G^{M}$, while the simple cuts in $G$ will be mapped to the other color of faces in $G^{M}$. The dual transformation on the original graph $G$ is equivalent to swapping colors on the medial graph.
\begin{center}
 \begin{tabular}{|c|c|c|}
    \hline
    Primal & Dual & Medial\\
    \hline \hline
    $Z(G)$        & $Z(G^{*})=C^{*}(G^{**})=C^{*}(G)$ & $Z(G^{M})=Z(G+G^{*})$ \\
    $C^{*}(G)$ & $C^{*}(G^{*})=Z(G)$                & $C^{*}(G^{M})$ \\   
    \hline
 \end{tabular}
\end{center}
         
\section{Non-planar surfaces}
 
In the last section we talked about planar graphs. In what follows we will discuss graphs embeddable in more general compact surfaces.
The sphere and plane have the same Euler characteristic and any graph embeddable in one is embeddable in the other. 

A graph embedded in a surface $\Sigma$ with faces that cover the entire surface is known as a tessellation of $\Sigma$.  

Euler's formula can be extended to closed, orientable surfaces
\begin{equation}
|F(G)|-|E(G)|+|V(G)|=\chi(\Sigma).
\end{equation}
 Where $\chi = 2 - 2g$ is known as the Euler characteristic for closed, orientable surfaces with $g$ handles.  A tessellation of a surface (2-manifold) is a 2-complex.  
 
An \emph{orientable} surface is a surface where a consistent normal vector can be defined. 

We can salvage most of the machinery from the last section by noting that the facial cycles, again with one redundancy, now generate most of the cycle space. The faces form a basis for all cycles that can be expressed as boundaries of some region of $S$. We will refer to any combination of facial cycles as boundary or homologically trivial cycles. Now, we have additional cycles that cannot be expressed as combinations of facial cycles. These cycles generate the remainder of the cycle space and are known as homologically non-trivial or boundary-less cycles. 

\begin{figure}[htb]
\center{\includegraphics[width=0.6\columnwidth]{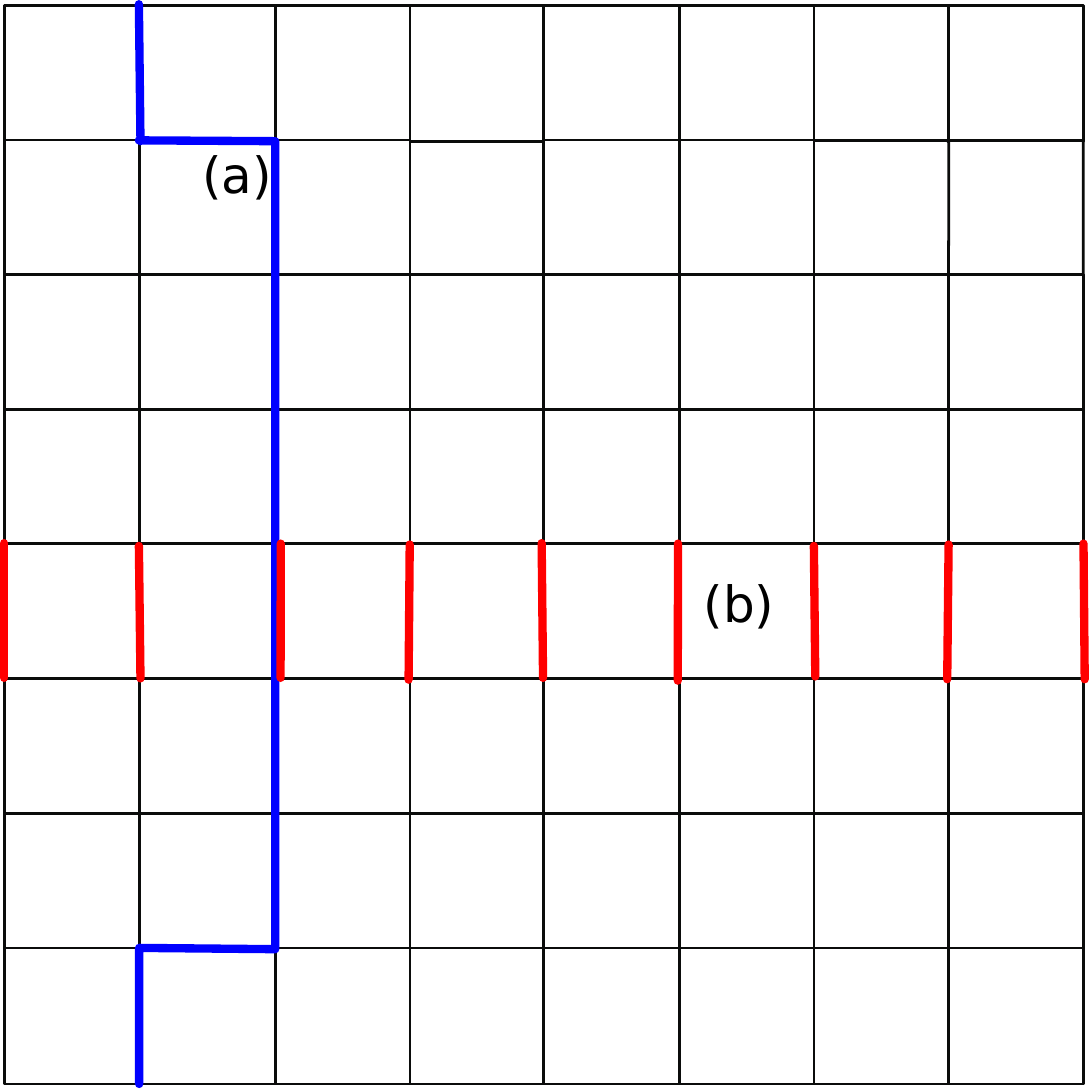}}
\caption{The graph above is drawn with periodic boundary conditions. (a) a non-trivial cycle. (b) a non-trivial cut. A non-trivial cycle(cut) cannot be expressed as a combination of simple cycles(cuts).} 
\end{figure}

\begin{figure}[htb]
\center{\includegraphics[width=0.75\columnwidth]{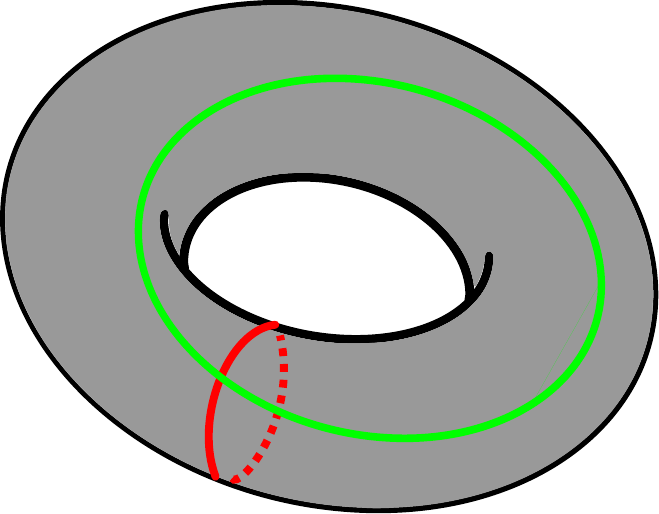}}
\caption{\label{fig:Boundaryless} The one-handled torus has two boundaryless cycles (red and green). Notice that these loops do not have the notion of bounding any area of the surface. They are also independent in that one cannot be deformed into the other. After a medial transformation on the primal graph, the non-simple cuts are mapped to boundaryless cycles.}
\end{figure}

We can define a group based on the loops that can reside on a surface. The elements are loops which are defined up to deformation. That is two loops that can be deformed into each other without breaking the loop correspond to the same element in the group. All boundary cycles are equivalent for a connected surface and correspond to the identity element. The remaining boundary-less cycles represent non-trivial elements of the group. The group operation can be represented visually as combining loops. We see that combining two boundary cycles always yields a boundary cycle and the combination of a boundary and boundary-less cycle is equivalent to the same boundary-less cycle. We will also define the combination of two equivalent boundary-less cycles to be the identity. In group-theoretic terms all elements of the group associated with loops on a surface must satisfy $g_{i}^{2}=I, \forall g_i\in G$  The group $G$ we have defined is known as the fundamental $Z_{2}$ homology group of the surface $\Sigma$. For the torus there are two boundary-less cycles (Fig.~\ref{fig:Boundaryless}). In general, for a $g$-handled orientable surface there are $2g$ boundary-less cycles. In the $Z_{2}$ homology the $2g$ boundary cycles generate the fundamental homology group of the surface. 

The dual tessellation corresponds to another graph embeddable in the same surface and, therefore, has the same Euler characteristic as the primal graph. The cut space will be modified in an analogous way as the cycle space. The simple cuts still correspond to the simple cycles of the dual graph. The primal and dual graphs will have the same $Z_2$ homology groups and the number of boundary-less cycles will equal the number of non-simple cuts. A non-simple cut is a cut that cannot be expressed as a combination of simple cuts.

The medial graph transformation also produces a tessellation of the surface with the same Euler characteristic. The new graph will be 4-valent and the same arguments apply as in the last section.   

The Euler characteristic of a surface $\Sigma$ can also be calculated by an alternating sum of Betti numbers:
\begin{equation}
\sum^{\infty}_{i=0}(-1)^{i}b_{i}.
\end{equation}

 Where $b_{0}$ is the number connected components of the surface, $b_{1}$ is the number of holes in the surface, $b_{2}$ is the number of 3D voids in the surface, etc. We will be dealing only with orientable surfaces which can be embedded in 3 or less dimensions, and will not need higher Betti numbers. 
 
 The $Z_{2}$ fundamental homology group $H_{Z_{2},1}(\Sigma)$ is the group of cycles on a surface $\Sigma$. Cycles with a boundary are represented as the identity in the fundamental homology group. Boundary-less cycles have a non-trivial representation in the fundamental group. Loops that can be deformed to each other are mapped to the same elements of the fundamental group.
 
 The fundamental group has an intimate connection with the first Betti number. $b_{1}$ determines the number of copies of $Z_{2}$ in the fundamental group. For example, the sphere has $b_{1}=0$ and $H_{Z_{2},1}=0$ or the trivial group. The circle has $b_{1}=1$ and $H_ {Z_{2},1}=Z_{2}$ while the torus has $b_{1}=2$ and $H_{Z_{2},1}=Z_{2}\times Z_{2}$. This relationship holds for any orientable surface. 
 
\section{Stabilizer Codes}  

In this section we will introduce qubit stabilizer codes where the Hilbert space is comprised of many physical qubits.

A stabilizer code is a quantum error-correcting code where all stabilizer generators $S_i$ are Pauli operations. Additionally, 
the stabilizer generators commute with each other $[S_i,S_j]=0\forall i,j$. The set of all states that are $+1$ eigenvectors for all combinations of stabilizer generators span a subspace of the Hilbert space. We refer to this subspace as the codespace.

A stabilizer code on $n$ qubits will be specified by its stabilizer generators. The stabilizer group specified by the stabilizer generators is an abelian subgroup of the $n$-qubit Pauli group.  The \emph{Pauli group} is the group represented by $n$-fold tensor
products of the Pauli matrices
\begin{align}
I := \begin{bmatrix}1&0\\0&1\end{bmatrix},
X := \begin{bmatrix}0&1\\1&0\end{bmatrix},
Y := \begin{bmatrix}0&\!\!\!-i\\i&\!\!\!\phantom{-}0\end{bmatrix},
Z := \begin{bmatrix}1&\!\!\!\phantom{-}0\\0&\!\!\!-1\end{bmatrix}
\end{align}
multiplied by overall phases of $\pm 1$ or $\pm i$.

The generators of the stabilizer group are the checks that must be measured in an error-correcting code. We will often refer
to the stabilizer generators as stabilizers though technically stabilizer refers to an element of the group
and not a generator. When the distinction matters we will specifically state which we are using. 

The $s$ stabilizer generators stabilize a $2^s$ dimension subspace of the $2^n$ dimensional Hilbert space. The remaining logical subspace is a $2^{n-s}$ dimensional 
subspace that encodes $k=n-s$ logical qubits. Each logical qubit will have corresponding logical operators. These logical operators
commute with all elements of the stabilizer group and act on a logical qubit as single qubit operator acts on a qubit.

We say that a stabilizer code is an $[\![n, k, d]\!]$ stabilizer code when it encodes $k$ logical qubits 
in $n$ physical qubits with distance $d$.  
We define the \emph{Pauli weight} of an element of the Pauli group
as the number of non-identity tensor factors it contains; the \emph{distance}
of a stabilizer code is the minimum Pauli weight of a logical operator. Logical operators 
take a code to an orthogonal state within the codespace. A
stabilizer code's distance therefore captures the size of the smallest
operation that will transform a state in the codespace to an orthogonal
state in the codespace. 

When $k=1$ the code encodes a single logical qubit. The logical qubit has a logical zero state defined as
\begin{equation}
|\overline{0}\rangle := 2^{-n/2} \prod_{i}(I + S_i)|0\rangle^{\otimes n}. \\
\end{equation}

For a single qubit the $X$ operator takes the $|0\>$ to $|1\>$. For a logical qubit the logical $X$ denoted $\overline{X}$ does the same. 

\begin{equation}
|\overline{1}\rangle := \overline{X} |\overline{0}\rangle,
\end{equation}

The logical operators allow one to control an encoded logical qubit just like a regular qubit. The logical operators will have the same type of interrelations as single qubit operators. For example, $\overline{X}\overline{Z}=-\overline{Z}\overline{X}$. For the codes discussed in this paper the logical operators will often have a representation as a string of single qubit operators on a graph. 

For qubit stabilizer codes the Hamiltonian can be expressed as
\begin{equation}
H=-\sum_{i=1}^{n-k}S_{i}.
\end{equation}
Where each stabilizer generator ($S_{i}$) is a local Hamiltonian that commutes with all other $S_{j}$'s. In the topological codes we consider each stabilizer generator is of bounded Pauli weight. 

The ground space of the Hamiltonian is the codespace and the number of degenerate ground states in the ground space is $2^{k}$ which is the same as the dimension of the logical space.

Subsystem stabilizer codes \cite{Poulin:2005a, Zanardi:2001a} are a generalization of stabilizer codes where the information is protected in a subsystem instead of a subspace. More recently, topological subsystem codes were introduced \cite{Bombin:2010a}. The family of codes introduced in this paper do not include subsystem codes. 
 
\section{Kitaev's toric code}
In this section we will introduce Kitaev's toric code (KTC) in its original form and then show that the medial graph transformation can be used to map all planar graphs on which KTC is defined to equivalent 4-valent graphs. 

KTC and by extension Kitaev's quantum double model is defined on any graph embeddable in a $g$ genus torus where qubits (qudits generally) are placed on edges (oriented edges in the quantum double model). The faces and vertices define pairwise commuting stabilizers. A face stabilizer corresponds to a tensor product of $Z$'s on each edge of the boundary of the face. A vertex stabilizer corresponds to a tensor product of $X$'s on each edge of the co-boundary of the vertex. In this paper,
we will define KTC in precise graph-theoretic terms. KTC can be defined for any simple graph embeddable in a surface, however, we will only allow simple planar graphs where each vertex is 3-valent or higher. 

We restrict the graphs to planar graphs or graphs embeddable in a sphere, because we would like to be able to increase the distance of the code without changing the surface that the graph can be embedded in. For example, the complete graph on $5$ vertices can be embedded in a torus without crossings but if we try to use this as a unit cell and tile the torus we will generally need to use a higher genus torus. For this reason non-planar graphs do not give us an extendible family of codes and we will restrict our attention to planar graphs. 

We also require that the vertices are 3-valent or higher for the following reasons: (1) Single-valent vertices correspond to a single qubit stabilizer and is therefore completely uncorrelated with any other qubit in the lattice. We can remove these vertices as they add nothing to the code. (2) Any two-valent vertex has a two qubit stabilizer and from a coding standpoint is equivalent to the same lattice with one of the two edges contracted. This is because the distance of the code, the number of logical operators, as well as other properties will be unaffected by contracting the edge. 

In this section, we will be considering KTC on orientable, compact, closed surfaces. Planar graphs and surfaces with punctures (``holes'') will be discussed later. 

We can use the medial graph transformation to take the KTC on graphs described above to simple 4-valent planar graphs. Each edge in a 3 valent or higher planar graph will be part of exactly 2 faces or simple cycles and this can be shown to imply that the medial graph is simple and 4-valent. If we allowed 1 and 2-valent vertices the medial graph would still be 4-valent, however, it would not be simple. In the medial graph version of KTC, qubits are on vertices and both types of stabilizers are now faces of the graph. In this new picture with qubits on the vertices we can see that the toric codes are precisely the 4-valent planar graphs. A 4-valent planar graph is always two face colorable and each color will correspond to a type of stabilizer. These 4-valent planar graphs can be embedded in any $g$-genus torus or the sphere. For each boundary-less cycle on the torus there will be a corresponding boundary-less cycle for each of the two colors. KTC on a $g$ genus surface will, therefore, have $2g$ logical qubits. 

Now, that we have shown that the toric codes are a strict subset of the planar graphs, we will spend the rest of the paper discussing the other planar graphs that can give rise to homological codes.  

\begin{figure}[htb]
\center{
  \subfigure{\includegraphics[width=0.30\columnwidth]{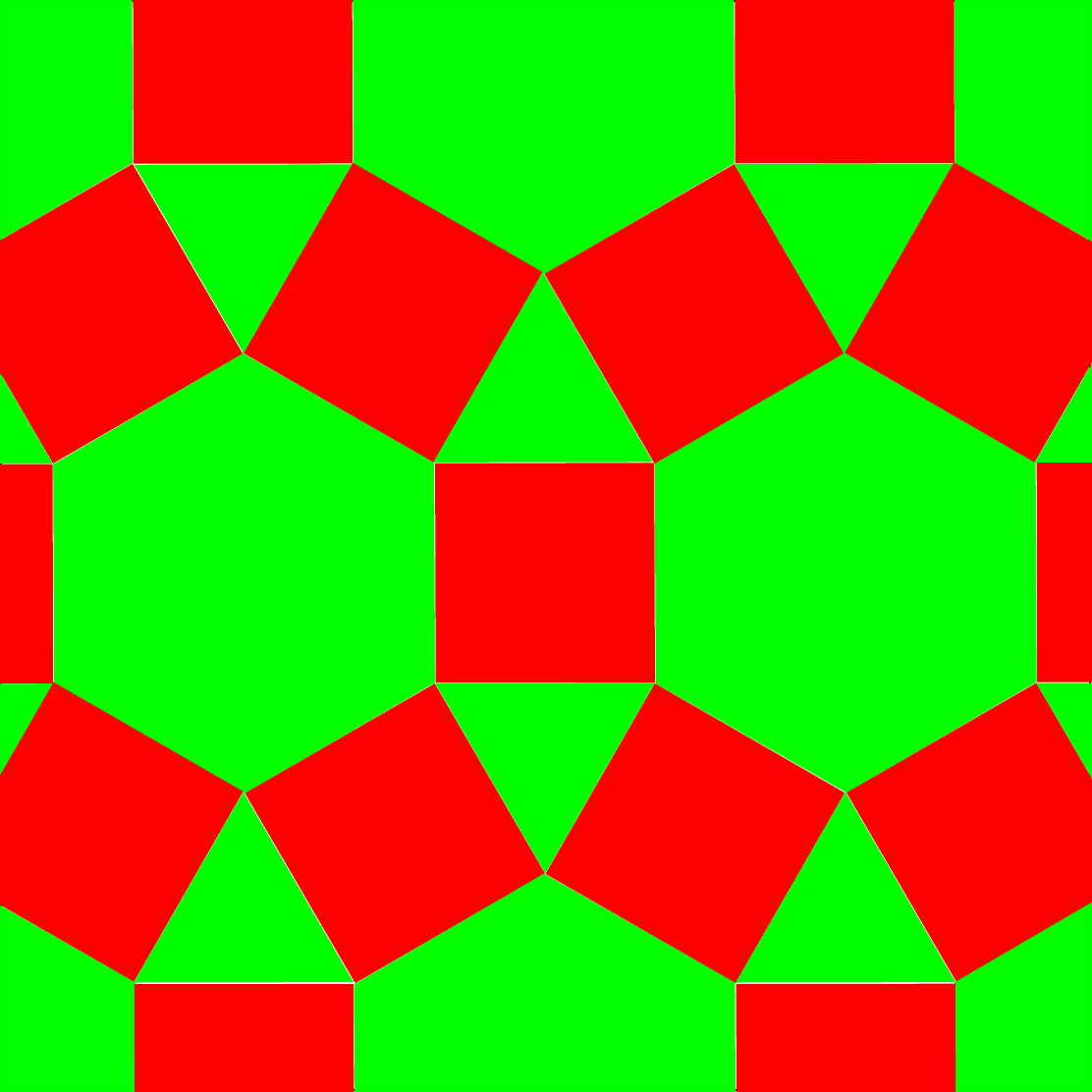}}
  \subfigure{\includegraphics[width=0.30\columnwidth]{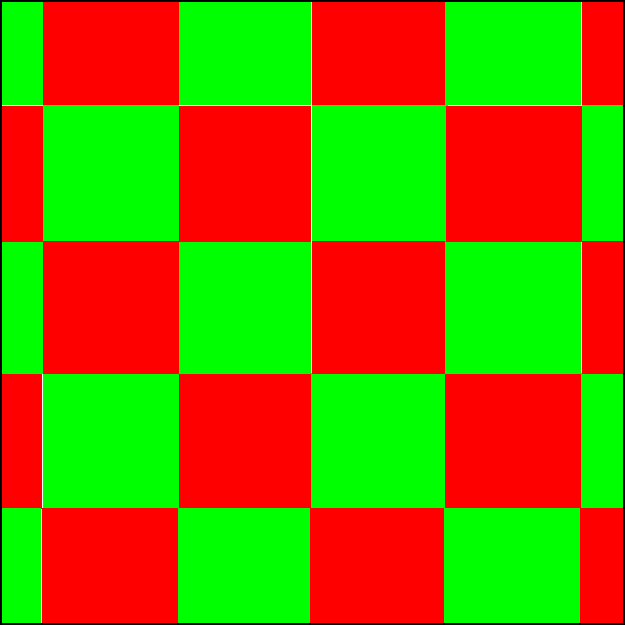}}
  \subfigure{\includegraphics[width=0.30\columnwidth]{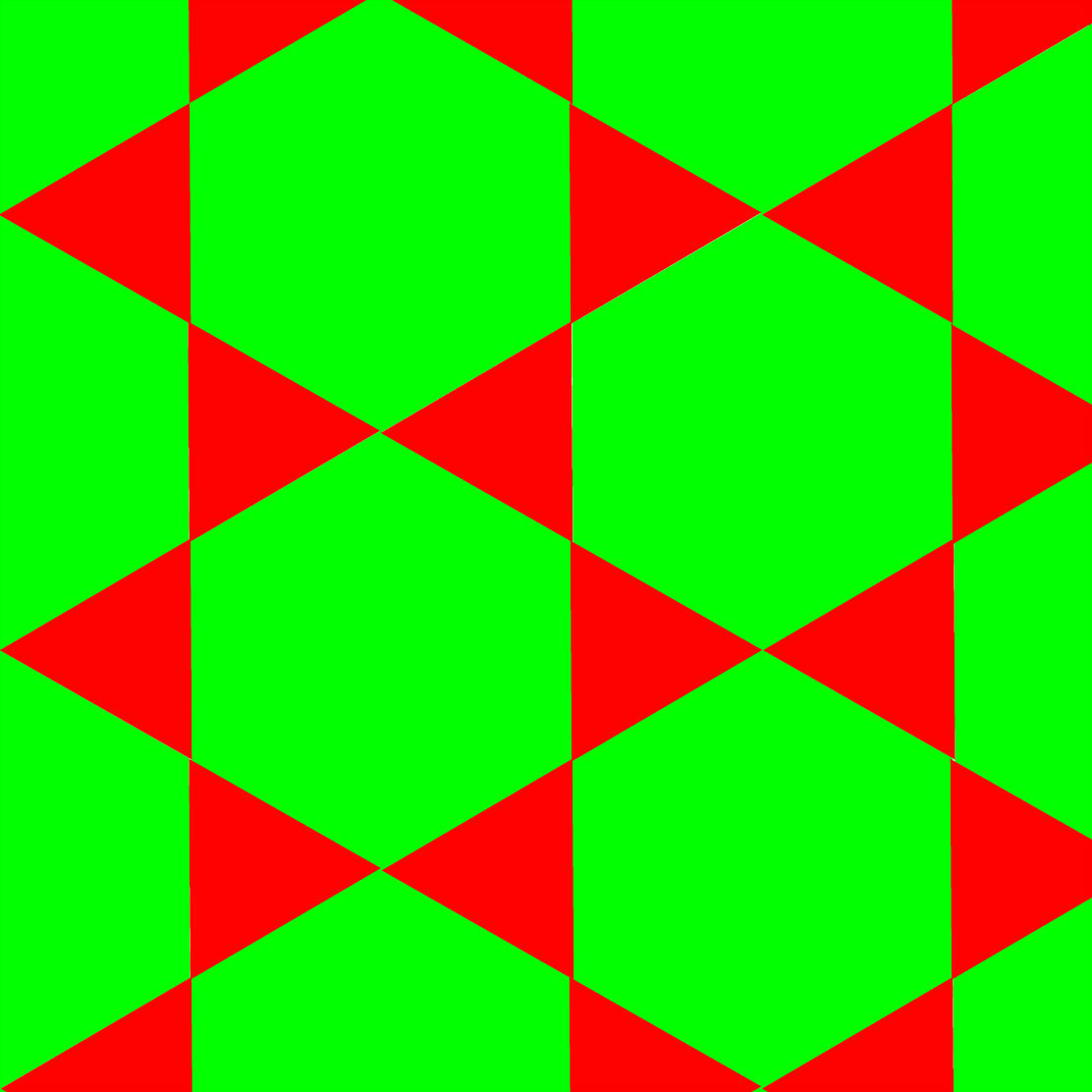}}
}
\caption{The $4$-valent vertex-regular graphs. All these graphs are two-colorable. Additionally, any 4-valent graph embedded in a surface is 2-colorable. These 4-valent graphs can all be used as toric codes by putting stabilizer generators on faces using one color for X-type stabilizers and the other color for Z-type stabilizers.}
\end{figure}

\vspace{1em}
\emph{Result 1: KTC can be defined on a graph $G$ iff $G$ is a 4-valent planar graph.}
\vspace{1em}

The colorability of the lattice will uniquely determine the types of quasiparticles. For concreteness I will put $X$-type stabilizers on the red faces and $Z$-type stabilizers on the green faces. A single qubit $X (Z)$ operator will anticommute with exactly two neighboring stabilizers on the green(red) faces. We can think of the faces as having quasiparticles on them. The absence of a quasiparticle is a trivial quasiparticles while single qubit $X$ or $Z$ operator will create a pair of red or green quasiparticles depending on the type of stabilizer that anticommutes with it. A red and green quasiparticle on neighboring faces is a special quasiparticle known as a dyon. We can calculate the statistics of these particles by exchanging two like particles and looking at the overall phase that the quantum state acquires. The trivial, red, and green quasiparticles pick up no phase when exchanged with a like quasiparticle and are therefore bosons. The dyon picks up a $-1$ phase under exchange and is a fermion. The quasiparticles can also have mutual statistics where exchanging different quasiparticle types can also cause the quantum state to pickup a phase. The three non-trivial quasiparticles have mutual fermion statistics. These particles are equivalent to the elements in the quantum double of $Z_{2}$. KTC is said to have $Z_{2}$ topological order.   

\section{Color Codes}

Another class of surface code introduced by Bombin et al. is the color code. 
The topological color codes (TCCs) are defined on 3-valent, 3-face-colorable planar graphs. These codes have two stabilizer generators (one $X$-type and one $Z$-type) per face with qubits on vertices. A 3-valent planar graph is 3-face-colorable iff all faces have an even number of vertices or equivalently the 3-valent graph is bipartite in vertices. In the TCCs the two stabilizer generators on each face commute because every face has an even number of vertices (qubits). In addition each face (stabilizer generator) shares $0 \bmod 2$ vertices (qubits) with neighboring faces. This ensures that all stabilizers commute and that all boundary cycles are of even weight. These 3-valent, 3-colorable graphs can also be used to tesselate a $g$-genus torus. Now, the boundary-less cycles will have a color and type ($X$ or $Z$). One color operator can be expressed as the product of the other two and we have a total of $4$ ($2$ colors $\times$ $2$ types) logical operators per boundary-less cycle. The TCCs as defined here encode $4g$ logical qubits in a $g$-genus torus. 

From the discussion in the last section we see that these two code families, KTC and TCCs, correspond to mutually exclusive families of planar graphs. 

In the following sections we will try to find other homological planar codes by looking at more general planar graphs with qubits on vertices.  

\begin{figure}[htb]
\center{
  \subfigure{\includegraphics[width=0.30\columnwidth]{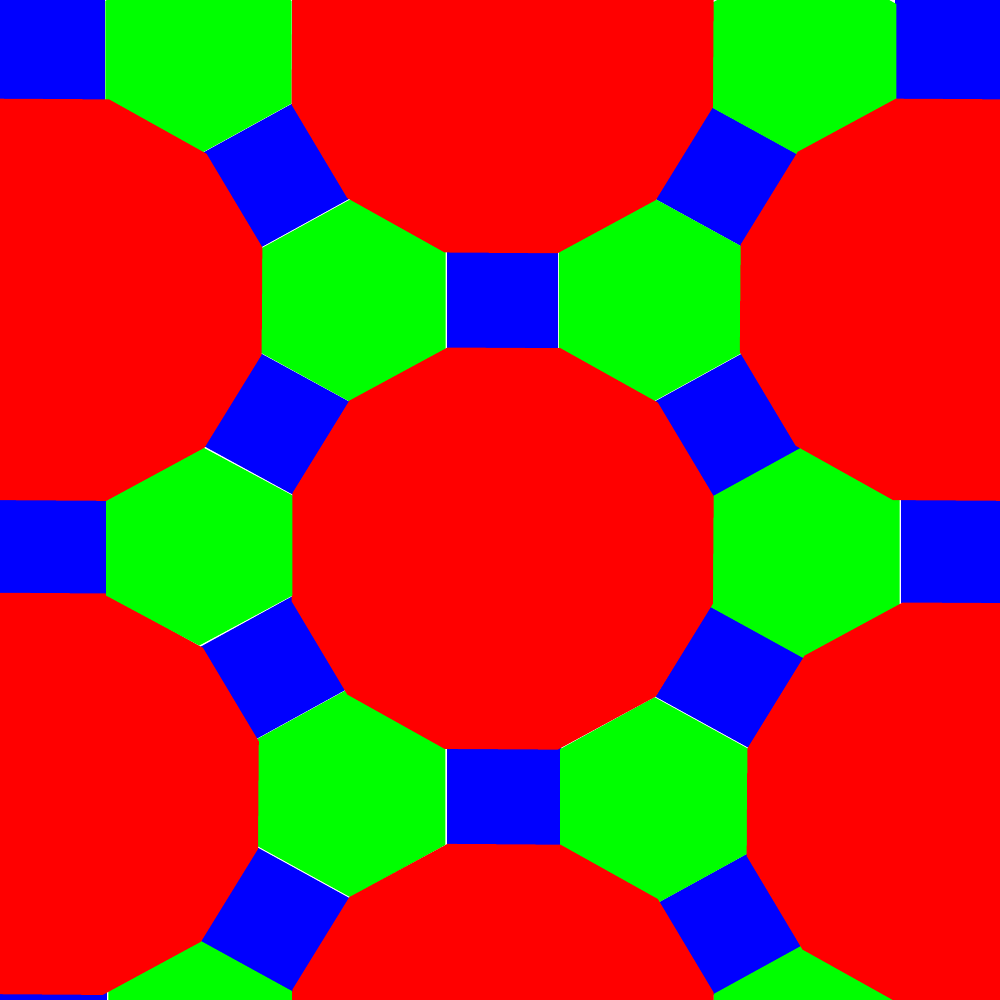}}
  \subfigure{\includegraphics[width=0.30\columnwidth]{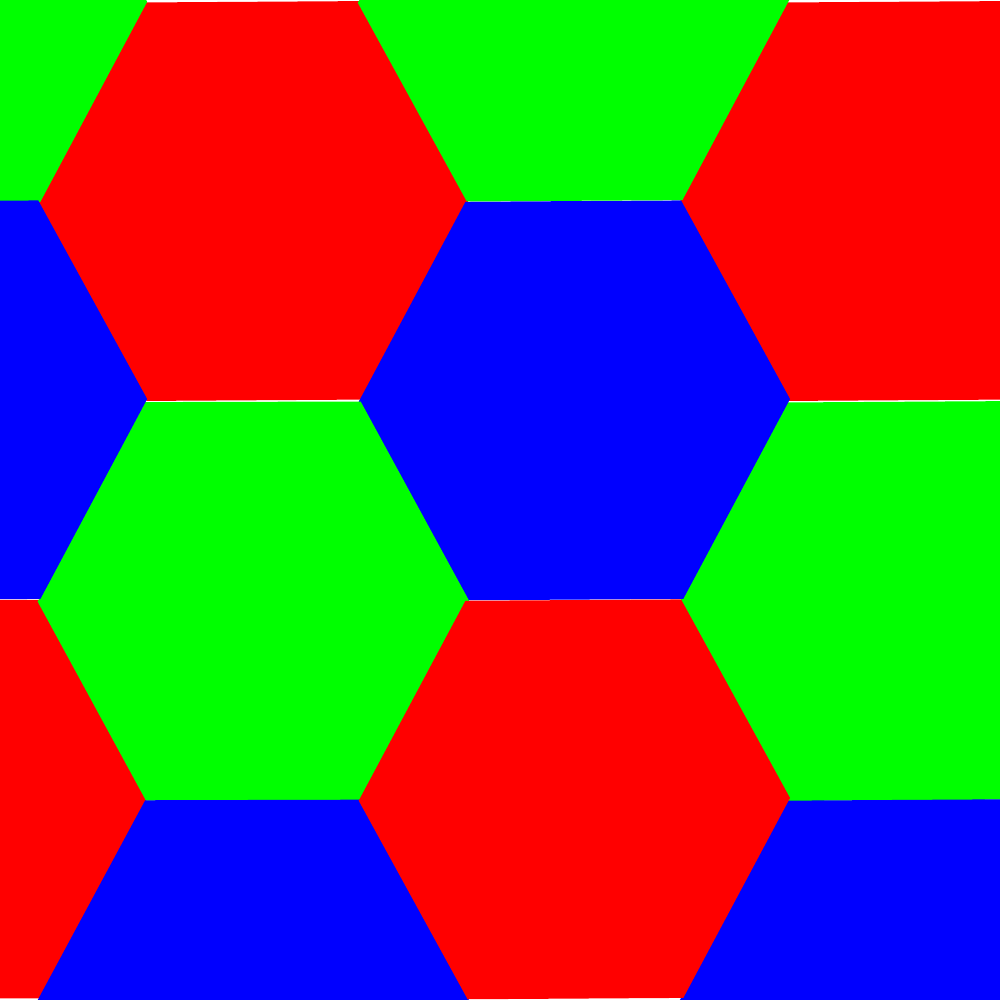}}
  \subfigure{\includegraphics[width=0.30\columnwidth]{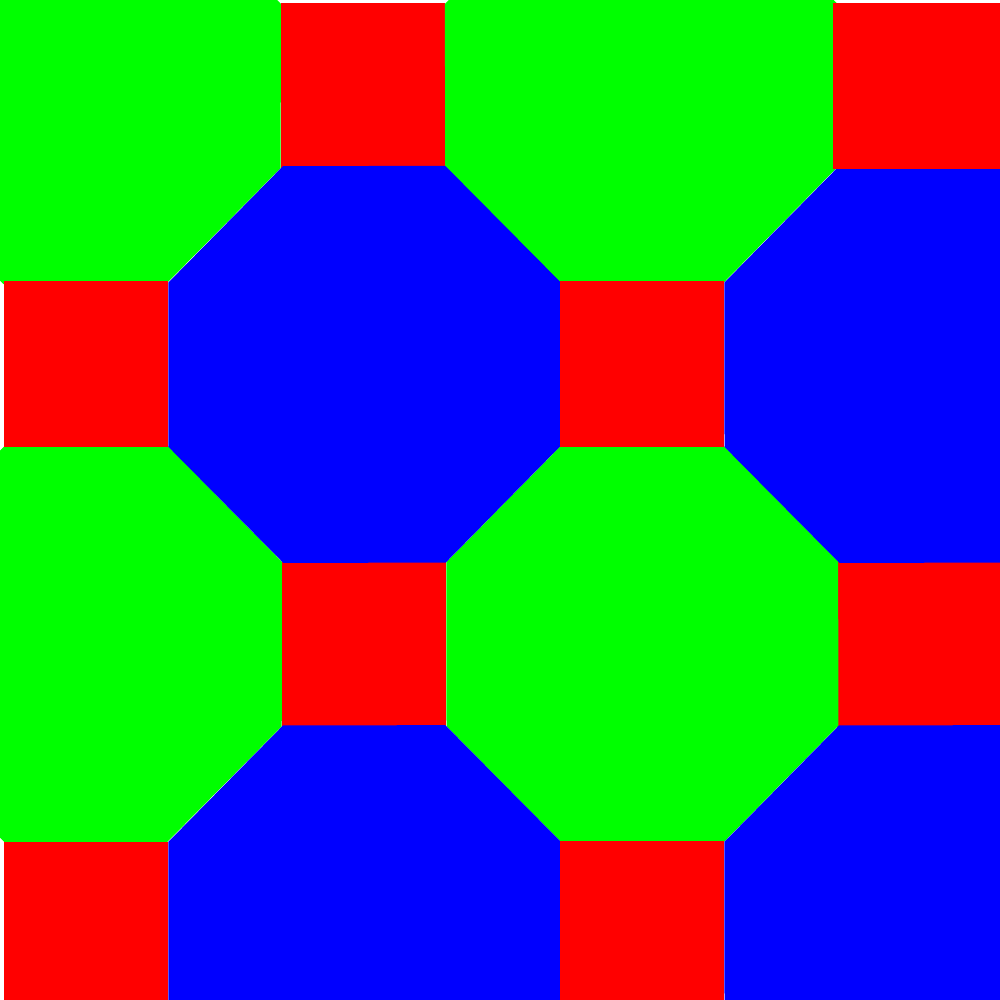}}
}
\caption{The 3-colorable, $3$-valent vertex-regular planar graphs. TCC are defined on these graphs by placing a stabilizer generator of each type on every face.}
\end{figure}

\vspace{1em}
\emph{Result 2: The TCCs can be defined on a graph $G$ iff $G$ is a planar 3-valent, 3-colorable graph.}
\vspace{1em}

Again, we can think of the quasiparticles as excitations on faces. Now, a quasiparticle can be one of three colors and each color can be either $X$ or $Z$-type. The 10 bosons are the trivial charge; a red, green, or blue, $X$ or $Z$-type excitation; and a $X$ and $Z$-type excitation on a red, green, or blue face. The rest of the quasiparticles are fermionic dyons. They consist of a $X (Z)$-type excitation on one color of face and a $Z (X)$-type excitation on a neighboring face of a different color. There are 6 ways to combine these excitations into distinct dyons and hence 6 fermions. In total there are 16 quasiparticles and they turn out to be equivalent to the quantum double of $Z_{2}\times Z_{2}$.    

\section{Homological Stabilizer Codes}
 
As we have seen in the last two sections, two distinct classes of graphs give rise to two different codes which have different types of quasiparticles. These graphs by no means exhaust the possible types of graphs embeddable in a surface or the plane for that matter. What other graphs can be used as homological codes? Can we find models with different types of quasiparticle excitations (topological orders) by using other classes of graphs?     

In this section, we will discuss more general stabilizer codes known as HSCs. We will construct these codes on planar graphs embeddable in a $g$ genus torus. Our goal is to encode logical information in the boundary-less cycles of these surfaces. We give a set of graph-theoretic conditions that determine which graphs can be used to construct HSCs.

First, we will define a HSC by specifying the properties of the graphs on which they are constructed. We will place qubits on vertices \footnote{All graphs with qubits on edges were already classified as toric codes and were mapped to 4-valent graphs by the medial graph transformation. Graphs with qubits on faces are simply the dual graph of the codes we are discussing so there is no loss of generality in this assumption.} and require that each face correspond precisely to a stabilizer of the code and the stabilizer has non-trivial support {\em i.e.} Pauli operators ($X$, $Y$, or $Z$) on each qubit (vertex) in that face. We have assigned stabilizers such that they are always boundary cycles and hence a trivial operation on the code will correspond to a homologically trivial loop on the surface. Furthermore, each stabilizer generator has support only on a particular face. This doesn't rule out putting multiple stabilizers generators on the same face, however, the product of these stabilizers must still be a boundary cycle. We impose this condition because the product of any two cycles in our graph must always give another cycle. This condition can be seen to allow two stabilizer generators on a face only when all elements anticommute. Since the two stabilizers must commute we conclude that any face with two or more stabilizer generators must be of even weight. If we try to add a third stabilizer generator to a face we must choose it such that it anticommutes with all single qubit elements in the other two stabilizer generators. There is only one such stabilizer that satisfies this condition and it is precisely the product of the other two generator and is therefore not independent. We conclude that at most two stabilizer generators can occupy a face. We can summarize these conditions as rule I.    

\vspace{1em}
I. {\em All stabilizer generators in a HSC have non-trivial support on qubits (vertices) that correspond precisely to the faces of a tessellation on some surface.}
\vspace{1em}

Later we will discuss the creation of punctures in the surface. Punctures can be thought of as faces without stabilizer support. 

The stabilizers form an abelian subgroup and therefore the product of any two stabilizers is always a stabilizer. Similarly, the product of any two facial cycles is always a facial cycle. Therefore, the product of any two stabilizers, as we have defined them, will be a facial cycle. 

We also require that the number of logical operators is independent of lattice size by requiring that the number of logical operators is a function of the boundary-less cycles of the surface. In other words, the number of logical operators is independent of lattice size. This condition is referred to as scale invariance in \cite{Yoshida:2011a}.

\vspace{1em}
II. {\em Logical operators in a HSC correspond to boundary-less cycles of the surface.}
\vspace{1em}

As a corollary to II we can immediately rule out all graphs with local logical operators. We see that at the very least a HSC must detect any single qubit error. We can express this property in terms of the stabilizer generators.

\vspace{1em}
IIA.{\em Every qubit in a HSC must be part of at least two stabilizer generators with different types of Pauli operators applied to that qubit.} 
\vspace{1em}

If this were not the case a weight one local logical operator would exist on that qubit.

We also require that at each vertex the same number and type of check operators are applied to the qubit at that vertex. This ``vertex regularity'' condition is a feature of all known homological codes. 

The excitations (violated stabilizers) caused by applying a Pauli operator to a qubit can be thought of as quasiparticles and the stabilizer type and number of stabilizers at each vertex will determine the fusion/splitting channels for the quasiparticles. If vertex regularity was not enforced the code would have fusion channels that differ by location in the lattice. This would change the excitation types at different areas of the surface and there would not be a consistent topological order. Recently, lattice defects similar to this have been studied and are referred to as twists. We will refer to the set of Pauli operators that check a vertex as its label set. The label set for a qubit is a list of all stabilizers that have non-trivial support at that qubit. The label set for a vertex contains the Pauli operators from incident faces. For each face we record the type of Pauli operator that has support on that qubit. For example, in the toric code the label set is $l=\{X,Z,X,Z\}$. When two checks are on a face we must express the label sets in pairs. Where a pair corresponds to the two different labels on a face. For example, in the color codes: $l=\{\{X,Z\}, \{X,Z\}, \{X,Z\}\}$. We can express the ``vertex regularity'' condition succinctly in the language of label sets.

\vspace{1em}
III. {\em Each qubit (vertex) in a HSC must have the same label set up to label set equivalencies (see below).}
\vspace{1em}

Rule II excludes single-valent vertices as the vertex cannot be part of a cycle. A connected graph where each vertex is two-valent is simply a single polygon. All vertices are part of a single face and have the same stabilizer generator(s). Polygons with an odd number of vertices can only have a single stabilizer generator and cannot protect against even a single error. Polygons with an even number of vertices have two stabilizer generators and label set $l=\{X,Z\}$ (up to equivalencies) at each vertex. These are the class of error-detecting codes with parameters: $[[2n,2n-2,2]]$. These codes are not homological as logical operators exist of weight 2, and we will not discuss them further. Therefore, we seek graphs of valency $3$ or greater.      

\begin{figure}[htb]
\center{\includegraphics[width=0.75\columnwidth]{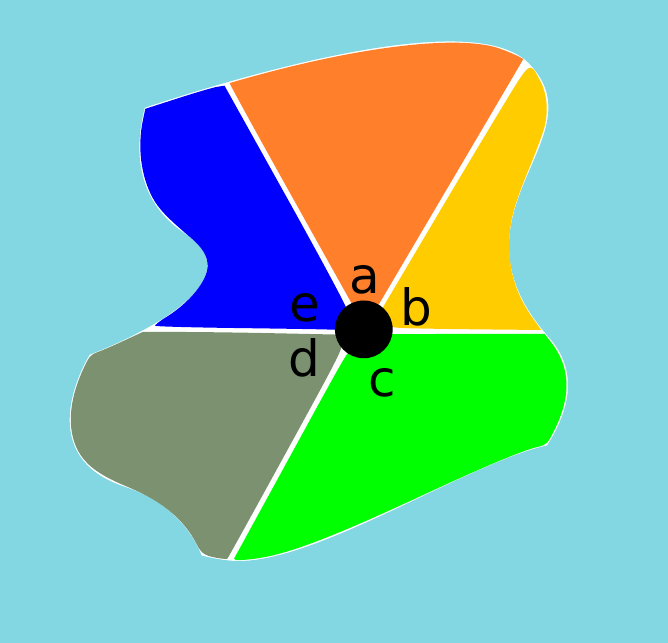}}
\caption{\label{fig:codebound} A 5-valent graph with a label on each face. The stabilizers (faces) cannot be made to commute. The vertex in the center is a qubit incident on 5 different faces. Each face corresponds to a stabilizer generator and each label on the face is a Pauli operator. If two faces only share a single vertex they must have the same label or else the corresponding stabilizers will not commute. Therefore we see that the following labels are equal $a=d=c$ and $b=d=e$. Which implies that all labels are equal. This is not allowed by IIA and therefore 5-valent and higher vertices are not consistent with rules I-III.}
\end{figure}

Rule I requires that stabilizer generators occupy faces. These stabilizers must commute with each other. Using Rules I and II we will show that the ramifications of this commutativity condition will greatly limit the allowed tessellations of a surface. If two faces overlap at a single vertex, the vertex will have one or more labels for each face. Lets call the label for one face a and the other b. The faces must have the same label type at that vertex (a=b), otherwise they do not commute. For a 5-valent lattice, the only way for the 5 incident faces to commute is for all 5 labels types to be equal. This creates a single weight logical operator and violates II. A proof in pictures is presented below (Fig.~\ref{fig:codebound}) that no $5$-valent or higher vertices can be present in a tessellation that satisfies rules I-III. 

\begin{figure}[htb]
\center{\includegraphics[width=0.75\columnwidth]{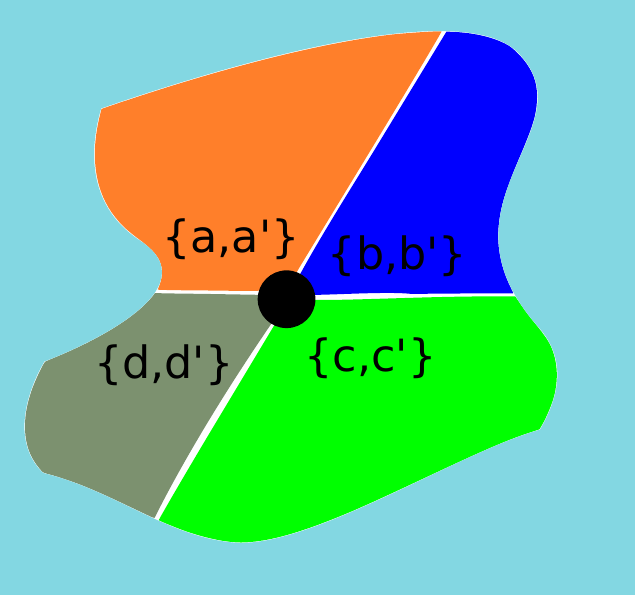}}
\caption{\label{fig:codebound2} A 4-valent graph with two labels per face. The stabilizers (faces) cannot be made to commute. The vertex in the center is incident on 4 different faces. The labels on a face must be different by I and if two faces share only a single vertex their labels must be equal. From this we have that $a=c=c'$ which is not allowed. Therefore, faces with two stabilizers can only be in graphs with 3-valent or less vertices.}
\end{figure}

A face that contains two stabilizer generators will have additional constraints. Each face below has two label which must be different and as with the single label case if two faces share only a single vertex the labels on these faces must be equal. As shown in Fig.~\ref{fig:codebound2} a graph with two stabilizers per face must be 3-valent or less. 

From these three rules we have reduced the allowed planar graphs to the subset of 3 and 4 valent planar graphs. 
Additionally, we have shown that if two stabilizer generators are present on faces then the graph is at most 3-valent.

\section{Label Set Equivalencies}

Now, we will discuss allowed transformations on the label set. These transformations produce equivalent codes in terms of distance, number of logical operators, number of stabilizers, and number and type of quasiparticles. When we discuss quasiparticles we can think of the label set transformations as local relabelings of quasiparticles. Imagine two observers at different locations on the lattice. They gives names to the types of quasiparticles and determine the statistics of the particles. The number of distinct types of quasiparticles and their exchange properties will otherwise be unchanged by a relabeling. We can use this label set equivalency to enlarge the classes of codes discussed above. We will show that the topological entanglement entropy and hence the topological order is unchanged by these transformations. 

The topological entanglement entropy (TEE) \cite{Kitaev:2005a, Levin:2006a} is a measure of the topological order of a system. The TEE has a correction to the area law scaling of the mutual information called the topological entropy ($\gamma$). The topological entropy can be expressed as $\gamma = \log D$. Where $D$ is the quantum dimension of the system and is the sum of the dimensions of the particles in the anyon model. 

First, any Pauli operator in the label set can be replaced by another Pauli operator as long as the number of labels of each type remain constant. The KTC label set can be transformed to other equivalent codes as follows: $\{X,Z,X,Z\} \rightarrow \{Y,Z,Y,Z\} \rightarrow \{Y,X,Y,X\}$. The label set transformation can preformed at any number of vertices. The transformation $\{X,Z,X,Z\} \rightarrow \{Z,Z,Z,Z\}$ is forbidden as the number of $Z$-type operators is different in each set. These transformations are just rotations by $\pi$ about the $X, Y$, or $Z$ axis on the Bloch sphere and, hence, can be performed by local Clifford unitaries (LC). Local unitaries cannot change the mutual information of a system and therefore the TEE and by extension the topological order is left unchanged by the transformation.

\begin{figure}[htb]
\center{\includegraphics[width=0.75\columnwidth]{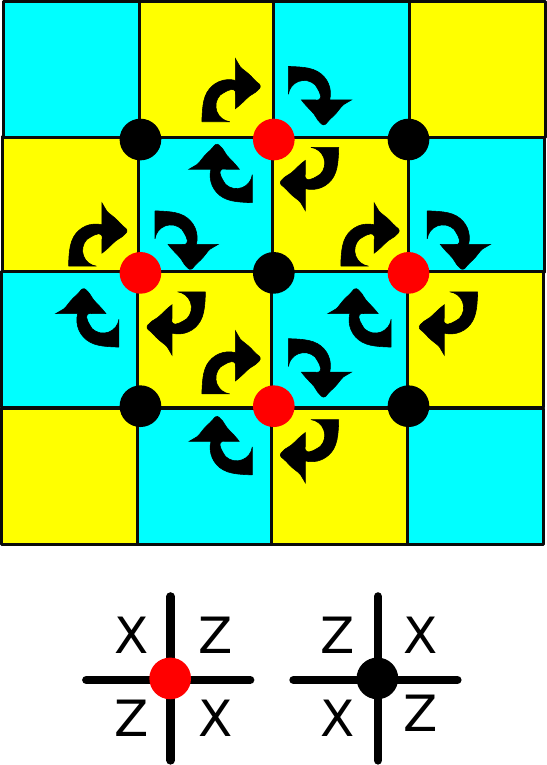}}
\caption{\label{fig:ktctolwpm} KTC shown with red and black vertices corresponding to the label sets $\{X,Z,X,Z\}$ and  $\{Z,X,Z,X\}$, respectively. We arbitrarily choose to start with the upper left label and proceed clockwise for both label sets. A set of $90^{\circ}$ rotations on the red vertices' label set, shown above, maps to KTC to the LWPM. The proves that the two models are label set equivalent.}
\end{figure}

Secondly, the label set may be cyclically permuted at any vertex. We express this as: $\{a,b,c,d\} \rightarrow \{d,a,b,c\}$. A cyclic permutation of a bipartition of vertices on the square lattice toric code can be shown to transform it to the equivalent LWPM see Fig.~\ref{fig:ktctolwpm}. A generic 4-valent graph has the label set $\{a,b,a,b\}$. Where the faces that share only one vertex must be given the same label to commute. This permutation does not change the ground state of the Hamiltonian and for stabilizer Hamiltonians the entire spectrum is unchanged. All properties of the underlying code are left unchanged by this operation. Additionally, the anyons in the model are unchanged by this permutation. They just appear as excitations on different faces. Thus, the topological order is unchanged by this operation. 

\begin{figure}[htb]
\center{\includegraphics[width=0.75\columnwidth]{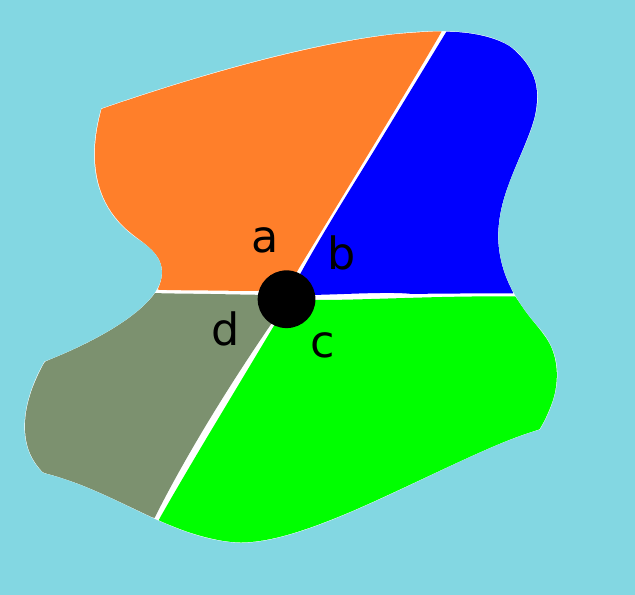}}
\caption{\label{fig:codebound3} Constraints on a general 4-valent graph with a label on each face. For the stabilizers to commute $a=c$ and $b=d$. From IIA $a\ne b$. Up to label set equivalencies this is the same label set as KTC.}
\end{figure}

All codes on 4-valent graphs are, up to label set equivalencies, of the form $\{a,b,a,b\}$. These are precisely the KTC. A proof in pictures is shown in Fig.~\ref{fig:codebound3}.

\vspace{1em}
\emph{Up to label set equivalencies KTC are the only stabilizer codes on $4$-valent graphs satisfying I-III.}
\vspace{1em}

The color codes which reside on 3-valent, 3-colorable lattices have two labels on each face. Again, we can replace Pauli operators with new Pauli operators. We can also cyclically permute the labels, but we must keep to pairs of labels together. Additionally, we can switch the order of all pairs. Some equivalent label set of the color codes are: $\{\{X,Z\}, \{X,Z\}, \{X,Z\}\}$, $\{\{X,Y\}, \{X,Y\}, \{X,Y\}\}$, $\{\{Z,Y\}, \{Z,Y\}, \{Z,Y\}\}$. 

By exhaustive counting we find that there are 3 distinct types of label sets. That is, up to label set equivalencies, we find the following label sets for $3$-colorable, $3$-valent lattices are: $\{\{X,Z\}, \{X,Z\}, \{X,Z\}\}, \{\{X,Z\}, \{X,Z\}, \{X,Y\}\}$, and $\{\{X,Z\}, \{X,Y\}, \{Z,Y\}\}$. These label sets change the number of stabilizers that anticommute with single Pauli operators. In other words, the number of anyons created from the vacuum differ among these label sets. The type and exchange statistics of the quasiparticles, however, will be the same as in the TCCs. The label sets equivalencies as we defined them are strictly local. The operations are on individual qubits. If we apply local unitary operations on a bounded number of qubits we can map the above label sets to each other. 
\begin{figure}[htb]
\center{\includegraphics[width=0.75\columnwidth]{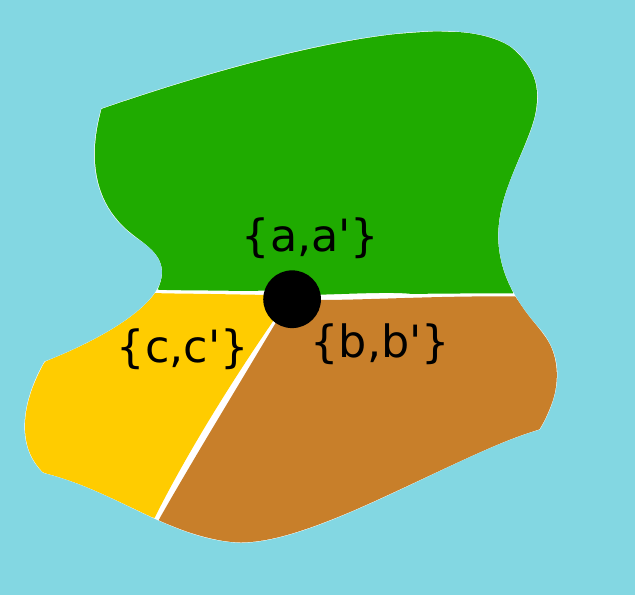}}
\caption{\label{fig:codebound4} Constraints on a general 3-valent graph with two labels per face. From I we have $a\ne a', b\ne b'$ and $c\ne c' $. Up to label set equivalencies there are 3 label sets that satisfy these constraints. One is the TCC, the other two are similar but not label set equivalent codes.}
\end{figure} 

\vspace{1em}
\emph{Up to label set equivalencies there are 3 classes of HSCs on $3$-valent, $3$-colorable graphs satisfying I-III. They all have similar topological properties as the TCCs.}
\vspace{1em}

It can be shown that any HSC satisfying I and IIA can only be
defined on graphs that are 4 valent or less. In fact, I and IIA alone restrict the allowed label sets on a 3-valent, 3-colorable graph to the 3 discussed above.
Similarly, I and IIA alone restrict the allowed label sets of a 4-valent graph to the one discussed above. 

Now, that we have given a concise list of properties defining HSCs as well as equivalency conditions, we will classify all 2D HSCs. 

\section{4-colorable graphs}

We have already discussed all the 4-valent and 3-valent, 3-colorable graphs. We have also shown that 5-valent and higher graphs will have a weight one logical operators and that $2$-valent graphs cannot produce topological codes. The only remaining class of graphs are the 3-valent, 4-colorable graphs. A 3-valent, 4-colorable graph must have an odd weight face. This odd weight face can only have a single stabilizer generator and hence the number of labels at each vertex bordering the odd weight face will be 5 or less. By III each vertex must have the same number of labels and therefore each vertex must border the same number of odd weight faces. As we will show below codes on 4-colorable graphs will have the number of logical operators scaling with lattice size. This violates II as the number of boundary-less cycles are independent of lattice size and therefore HSCs cannot be defined on 4-colorable graphs. 

A necessary condition for a code to encode logical information in the boundary-less cycles of a surface (II) is that the number of logical qubits stays constant as the surface size increases or equivalently as the tessellation becomes finer. In coding terms we demand that the number of logical qubits is a constant independent of the number of qubits. We can express this quantitatively as

\begin{equation}
\lim_{\mbox{qubits} \to\infty} \frac{\mbox{logical qubits}}{\mbox{qubits}}=0.
\end{equation} 

This equation demands that the density of logical qubits goes to zero in the limit of infinite lattice size.

In what follows we will show that all 3-valent graphs must have two labels per face to satisfy rule II. The two label condition on 3-valent graphs will then imply that 4-colorable graphs do not admit HSCs satisfying  I-III. 

For a 3-valent graph $3|V|=2|E|=3|F_{3}|+4|F_{4}|+\hdots$. Now, 
\begin{equation}
|V|=\frac{|F|\times F_{avg}}{3}.
\end{equation}

From the Euler characteristic we have $|F|-|E|+|V|=2-2g$. Using the fact that the graph is 3-valent we arrive at $|F|=|V|/2 + 2 -2g$. We can combine these equations to obtain a formula for $F_{avg}$:

\begin{equation*}
F_{avg}=\frac{3|V|}{2-2g+|V|/2}
\end{equation*} 

For fixed $g$, $F_{avg}$ will approach $6$ in the limit of large $|V|$. We now have an asymptotic value for $F_{avg}$. 

We can express (11) using only graph theoretic terms. 

\begin{equation}
\begin{split}
      \lim_{qubits \rightarrow \infty} \frac{\mbox{logical qubits}}{\mbox{qubits}}&=\frac{\mbox{qubits - stabilizer generators}}{\mbox{qubits}}\\
                  &=\frac{|V|-m |F|}{|V|}\rightarrow0.
\end{split}
\end{equation}

Where $m$ is the average number of stabilizer generators per face. We can express $|V|$ in terms of $|F|$.

\begin{equation}
\frac{|V|-m |F|}{|V|}=1-\frac{m|F|}{|F| F_{avg}/3}=1-\frac{3m}{F_{avg}}.
\end{equation} 

Now, we take the limit as $|V|$ goes to infinity and use our asymptotic value for $F_{avg}$. 
\begin{equation}
1-\frac{3m}{F_{avg}} \rightarrow 1-\frac{3m}{6}
\end{equation}

We see that $m$ must equal 2 for the density of logical operators to go to zero in the infinite lattice limit. In other words, each face of a 3-valent graph must have 2 stabilizer generators as no face can have more than 2 stabilizer generators. We can exclude all 3-valent graphs with odd-weight faces incident since these faces can have at most one stabilizer generator. Also, since all 3-valent graphs with even weight faces are three-colorable we have shown that the colorability of a 3-valent graph determines the presence of local logical operators.

We have now excluded all 5-valent or higher graphs as well as 3-valent, 4-colorable graphs as candidates for HSCs. We have also classified all label sets corresponding to HSCs for the remaining planar graphs.  

\section{Optimal HSCs}
The HSCs have code distance that grows with lattice size. KTC has $2g$ logical operators while the TCCs along with its variants discussed in section VIII. have $4g$ logical operators. Where $g$ is the genus of the surface that the code is embedded in. It was shown in \cite{Bravyi:2009a} that 2D stabilizer codes cannot scale better than: $kd^{2}=O(n)$. Where $k$, $d$, and $n$ are the number of logical qubits, code distance, and number of physical qubits, respectively. KTC and TCCs are examples of codes that achieve this bound. In fact, all HSCs achieve this bound. In this section we find the minimum weight stabilizer generators among the HSCs. 

In terms of graphs, we look for the smallest number of vertices per face in all graphs where HSCs are defined. Generally, faces have different numbers of vertices and we must minimize the average number of vertices per face ($F_{avg}$). In what follows we show that HSCs must have $F_{avg}\ge 4$. This implies that the square lattice KTC is optimal in terms of stabilizer weight. 

In a recent paper \cite{Aharonov:2011a} it is shown that the problem of \textsc{local commuting hamiltonian} for 3-body, qubit Hamiltonians is in the complexity class NP. The onset of topological order is stated as fundamental problem with extending their construction to 4-body, qubit Hamiltonians. They also prove that 3-body commuting qubit Hamiltonians cannot have topologically-ordered ground states. The HSCs discussed in this section have topological order and consist of local $n$-body commuting Hamiltonians. Where $n\ge 3$. While we can construct local commuting Hamiltonians with topological order and 3-body operators we must also include higher body operators. 

An obvious question is whether a HSC, which has topological order by design, can be constructed with $(n_{avg}<4)$-body local Hamiltonians. Each commuting local Hamiltonian will be a stabilizer generator in the HSC. We show that while HSCs with some 3-body operators can be constructed the average size of the local Hamiltonian is 4 or greater.

To prove this we show that the average number of vertices per face (stabilizer weight) is $\ge 4$. In this sense KTC on the square lattice is optimal. 

We will first discuss the 4-valent lattice case then we will briefly discuss the 3-valent cases. For 4-valent HSCs we can relate the number of edges to the number of vertices as
\begin{equation*}
2|E|=4|V|
\end{equation*}

and to the number of faces as
\begin{equation*}
2|E|=3|F_3|+4|F_{4}|+5|F_{6}|+\hdots = F_{avg}|F|.
\end{equation*}

Combining we get
\begin{equation*}
F_{avg}=4|V_4|/|F|
\end{equation*}

Using the Euler characteristic of a genus $g$ surface and the above equations we see that
\begin{equation*}
F_{avg}=4|V_4|/(2-2g+|V_4|).
\end{equation*}

$F_{avg}\approx 4$ when $g<<|V_4|$ and approaches $4$ quickly in the large lattice limit. 

A similar calculation shows that for 3-valent HSCs $F_{avg}\approx 6$. We see that while HSCs can be defined with some weight 3 stabilizer generators 
the average stabilizer size will be at least 4. Thus, a necessary condition for Hamiltonians corresponding to HSCs is that the Hamiltonian is $k$-local. Where $k_{avg}\ge4$. For general Hamiltonians the problem is still open. Namely, \emph{what is the minimum $k$ such that a $k_{avg}$-body commuting qubit Hamiltonian's ground state exhibits topological order?} 

The remainder of the paper discusses well-known features of topological codes in the language of graph theory.

\section{Punctures}

All the HSCs have stabilizer generators on faces which are part of some tessellation of a surface. Every connected, orientable surface has a number of boundary-less operators determined by the genus of the surface. This fixes the number of logical operators for a code. We can, however, puncture the surface by creating holes in it. A puncture can be seen as a relaxation of rule I. The first hole in a surface will not change the fundamental homology group and therefore the number of logical qubits will not change. To see this imagine putting a hole in a sphere, then make a loop around the hole. At first sight it may look like the loop has no boundary, however, the rest of the sphere provides the boundary and the loop is indeed a boundary cycle. Similar reasoning can be applied to any connected, orientable surface. Adding a second hole will change the fundamental homology group. In fact, each additional hole after the first will add a boundary-less cycle to the surface. We can think of the hole-type logical qubit as being encoded in two holes of the same type and color. There is a loop operator of the same type as the hole around either of the holes. There is also a path operator connecting the two holes of the opposite type. A pair of loop and path operators will anticommute and correspond to logical $X$ and $Z$ for the logical hole-type qubit. 

To add a hole to any of our homological stabilizer codes we remove a stabilizer generator from the code. The face that corresponds to this stabilizer generator will appear as a hole in the surface. The hole-type logical operators will be defined by color and type. In the 4-valent graphs, holes can be classified as one of two colors (red or green). The number of hole logical qubits in these surfaces will be

\begin{equation}
l_{c}  = h_{c} -1 \mbox{ , }c\in\{\mbox{red, green}\}. 
\end{equation} 

Where $l$ is the number of logical qubits and $h$ is the number of holes for a particular color $c$. 

In the 3-colorable, 3-valent graphs the holes will have one of three colors (red, green, or blue) and additionally a type($X$ or $Z$). A single face can now have one or two ``holes'' depending on whether one or both types of stabilizers stop being enforced at that face. We have a similar relation for the number of hole-type logical qubits for these graphs.

\begin{equation}
l^{i}_{c}  = h^{i}_{c} -1 \mbox{ , }c\in\{\mbox{red, green, blue}\}, i\in\{X, Z\}. 
\end{equation} 

Where $l$ is the number of logical qubits and $h$ is the number of holes for a particular color $c$ and Pauli type $i$. 

\section{Boundaries}

The holes in the last section can be thought of as internal boundaries, however, in this section we will consider external boundaries. These boundaries allow us to implement HSCs in the plane without the difficulties of constructing multi-genus tori. Boundaries will be defined by the type of string operator that can end on it. The types of possible boundaries for the HSCs, like holes, will be determined by the colorability of the graph. 

The interface between boundaries in this section will correspond to a change in color of the boundary. At the vertex on the interface rule III is relaxed, and the degree of this vertex is one less than the other bulk vertices.  

For the 2-colorable lattices there are only two types of boundary. We will assign each boundary a color based on the type of string that can end there. The changes in boundary-type will be marked by a 3-valent vertex while all other vertices will be 4-valent. To make larger boundaries of the same color we need to add weight two stabilizers to the outer edge. These weight two operators keep the outer color the same for the length of the edge and only allows one color of string to end on that edge. 

\begin{figure}[htb]
\center{
  \subfigure{\includegraphics[width=0.55\columnwidth]{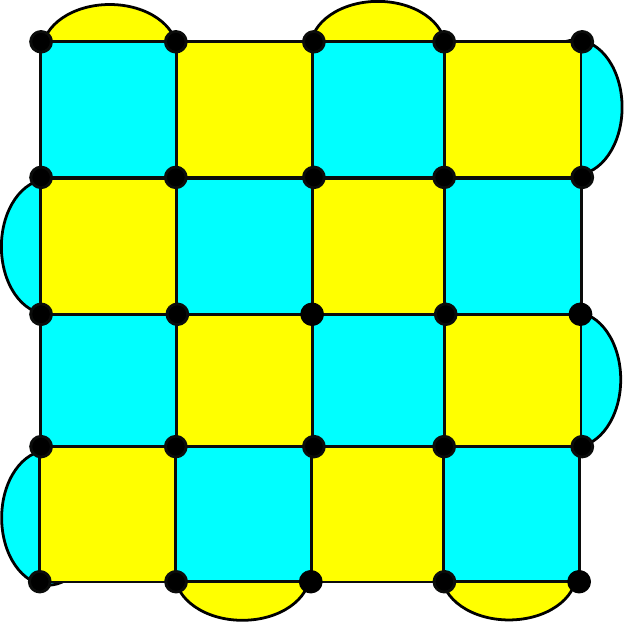}}
  \subfigure{\includegraphics[width=0.35\columnwidth]{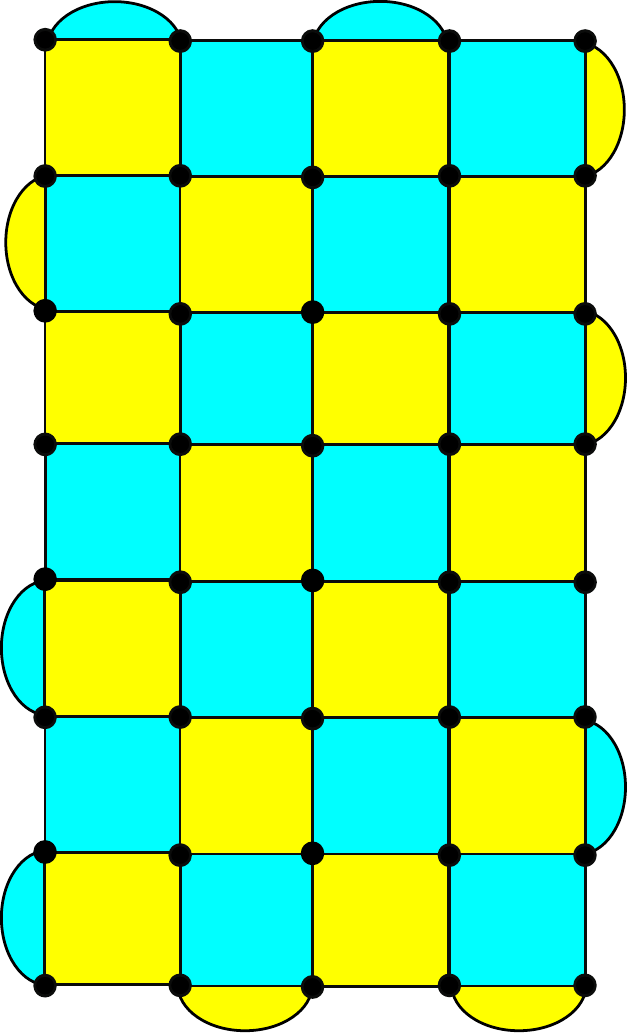}}
}
\caption{KTC in the plane. The weight two stabilizer generators must be added to give each side a consistent color. The KTC with 4 distinct colored boundaries encodes one logical qubit. The KTC with 6 distinct colored boundaries encodes 2 logical qubits. The distance of the code will be the weight of a string that connects like colored boundaries. The distance would be trivial without the introduction of the weight two stabilizers. We can also determine the number of logical qubits by counting the number of 3-valent vertices.}
\end{figure}

As shown by Bravyi and Kitaev \cite{Bravyi:1998a} the number of logical qubits is related to the number of distinct boundaries. The number of logical qubits = (distinct boundaries/2)-1. The total number of boundaries is always even as the two boundary types must alternate an even number of times as one goes along the outer boundary of the lattice. 

We can also determine the number of logical qubits in a code by counting the number of 3-valent vertices.

\begin{equation}
\mbox{logical qubits} = \frac{|V_3|}{2}-1
\end{equation}
 
For the 3-valent, 3-colorable lattice we have 3 types of boundaries. Again, we will label a boundary by the color of strings that can end there. A boundary will allow both $X$ and $Z$ strings to end at it. There will, therefore, be only three types of boundaries. Each boundary will be distinguished by a 2-valent vertex and all other vertices will be 3-valent.  Boundaries can alternate: red, blue, green as in the surface version of the color code or alternate between only two colors. If only two color are used as boundaries, a string operator of the third color can end by splitting into the other two colors and then ending on the appropriate colored boundaries. The number of logical operators will be the number of distinct boundaries minus 2. In terms of 2-valent vertices we have:
\begin{equation}
\mbox{logical qubits} = |V_2|-2.
\end{equation}  

\begin{figure}[htb]
\center{\includegraphics[width=0.75\columnwidth]{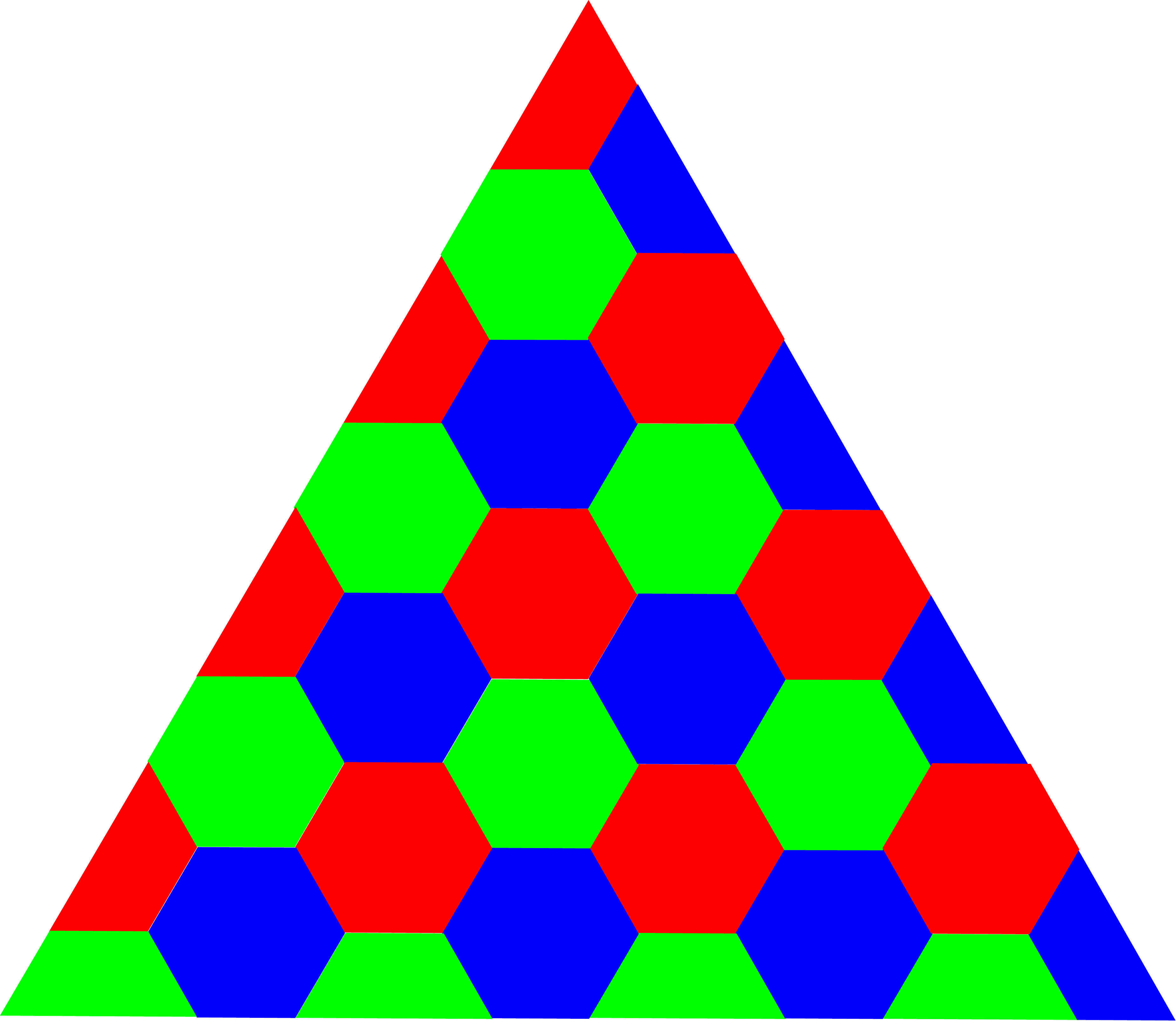}}
\caption{The 3-valent, 3-colorable 6.6.6 lattice. The boundaries are assigned a color based on what type of string can end at it. The right side with alternating blue and red faces allows only green strings to end there and we, therefore, call this boundary green.}
\end{figure}  

If one wants to add holes to either of the lattice types with boundary discussed above, they can do so but removing any stabilizer that does not border the external boundary. For lattices in the plane the addition of a single hole changes the fundamental homology group and the equations from the previous sections can be amended by adding one to the number of logical operators.  

 \section{Twists}
 
Bombin introduced the twist \cite{Bombin:2010a, Bombin:2010b} as a way for quasiparticles to change type. In the region of the lattice near a twist a quasiparticle can exist as a mixture of quasiparticle types. In some cases this mixture can be used as a composite quasiparticle with new statistics. In KTC the statistics of these new composite quasiparticles change the topological order from $Z_2$ to that of the computationally more powerful Ising anyon model. 

We can also describe the twist in graph theoretic terms. The vertex regularity condition (III.) can be relaxed at a small number of vertices that is independent of lattice size. In some cases this gives rise to a defect known as a twist. In the 4-valent lattices a twist is created by adding a 3-valent vertex. Stabilizers need to be changed throughout the lattice to accommodate the change, however, the vertex regularity condition is only broken at the twist. This 3-valent vertex necessarily makes the graph 3-colorable. 

In the 3-valent, 3-colorable lattices a twist is induced by adding an odd-weight face. The lattice will necessarily be 4-colorable. The lattice must be modified globally, however, the vertex regularity condition is still satisfied away from the twist. The quasiparticles can change color in this region. 

While this section in no way exhausts the possibilities for twists, it gives a graph theoretic picture of a twist based solely on colorability. 

Twists in 3-valent lattices get around the proof of section IX. because they occur at a number of sites that is constant with lattice size and in the large lattice limit do not change the number of logical qubits. 

\section{Conclusion}

We have seen that the KTC and TCC can be constructed on two distinct lattices. These lattices can be distinguished by their colorability alone. The colorability of a planar graph was shown to be the key feature in defining which HSCs were allowed on that graph. We showed that HSCs can only be defined on 4-valent or less lattices. Additionally, 3-valent 4-colorable lattices will necessarily have local logical operators and cannot be used for HSCs. Then, we showed that up to label set equivalencies KTC and the TCCs are the only HSCs on surfaces. We also discussed the notion of holes, boundaries, and twists and showed that they can also be interpreted in graph theoretical ways. We hope this graph-based description provides insight into some of the most promising ideas in quantum error correction and fault-tolerance. Also, the codes introduced in this paper were introduced constructively meaning that the graph that defines a HSC also provides a planar architecture for the code.

%

\begin{acknowledgments}
The author would like to acknowledge many fruitful discussions with Courtney Brell, Chris Cesare, Guillaume Duclos-Cianci, Andrew Landahl, and Gregory Crosswhite.
JTA was supported
in part by the National Science Foundation through Grant 0829944.JTA was supported in part by the
Laboratory Directed Research and Development program at Sandia National
Laboratories.  Sandia National Laboratories is a multi-program laboratory
managed and operated by Sandia Corporation, a wholly owned subsidiary of
Lockheed Martin Corporation, for the U.S.  Department of Energy's  National
Nuclear Security Administration under contract DE-AC04-94AL85000.
\end{acknowledgments}

%
\begin{appendix}
\section{Relaxing Rule III.}
Rule III. was introduced to make the types of quasiparticle excitations consistent throughout the lattice. In our discussion of twists we relaxed this rule at a constant number of locations on the lattice. From a stabilizer code standpoint, however, this rule may seem arbitrary. In this section we show that graphs combining 3 and 4-valent vertices in any regular way are not useful as HSCs. Therefore, rule III. is only necessary to ensure quasiparticle consistency. 
 
Using only rules I and II we have already discussed how 5-valent and higher graphs will always have local logical operators. Also, it can easily be shown that any face with at least one 4-valent vertex can have at most one stabilizer generator. 

For this section we imagine some type of lattice that combines 3 and 4-valent vertices in a semi-regular pattern. First, we will assume the number of each type of vertex (3 and 4-valent) goes to infinity in the large lattice limit. If this was not the case, we could treat the finite type of vertex as twists which we have already discussed. Since the number of faces $|F|$ in the lattice must also go to infinity in the large lattice limit we can say that $|V_{3}|,|V_{4}|\propto |F|$. We will also assume that the faces containing only 3-valent vertices have an even number of vertices. If we did not assume this we could use the arguments of section IX. to prove the logical qubits scale with lattice size. 
 
We can bound the number of vertices per face ($F_{avg}$) using the 4-valent and 3-valent, 3-colorable cases as lower and upper bounds, respectively.

\begin{equation}
4 < F_{avg} < 6
\end{equation}

We also require as in section IX. that the number of logical qubits is independent of lattice size:

\begin{equation}
\begin{split}
       \lim_{qubits \rightarrow \infty} \frac{\mbox{logical qubits}}{\mbox{qubits}}&=\frac{\mbox{qubits - stabilizer generators}}{\mbox{qubits}}\\
                  &=\frac{|V|-m |F|}{|V|}\rightarrow0.
\end{split}
\end{equation}

We can express $|V|$ as:
\begin{equation}
|V|=|V_{3}|+|V_{4}|=\frac{|F| F_{avg} + |V_{3}|}{4}.
\end{equation}

Now using (A2),

\begin{equation}
\begin{split}
1-\frac{m|F|}{|V|}=1-\frac{m|F|}{\frac{|F| F_{avg} + |V_{3}|}{4}} &=1-\frac{4m|F|}{|F|F_{avg}+|V_{3}|}\\
                &=1-\frac{4m}{F_{avg}+c}.
\end{split}
\end{equation}

Where $|V_{3}|=c|F|$, $0<c$ by the discussion above. 

For nonzero $c$ in the large lattice limit,
\begin{equation}
1-\frac{4m}{F_{avg}+c}\rightarrow 0.
\end{equation}

But, $4<F_{avg}+c$, implying that $m>1$. 

We have shown that the average number of stabilizer generators per face for a lattice with 3 and 4-valent vertices is greater than 1. 

If we assume some structure to our graphs. For example, each face has at least one 4-valent vertex. The average number of stabilizer generators per face is greater than one and as we discussed this can only be true for faces having only 3-valent vertices, which we assumed was not the case. We have now ruled out all semi-regular tilings of the plane that combine 3 and 4-valent vertices as candidate graphs for HSCs. 

We imagine the most favorable combination of 3 and 4-valent vertices would be to have a large contiguous region of the lattice of one type and the rest of the other. All faces with 4-valent vertices will have a single stabilizer generator while all faces consisting entirely of 3-valent vertices will have two stabilizer generators. Only at the border between regions will the faces with 3-valent vertices not be able to have two stabilizer generators. Even in this case when the border grows with lattice size, implicitly assumed when $|V_{3}|,|V_{4}|\propto |F|$, the number of local logical qubits grows with the size of the border. This example is also not useful for finding new codes, as the two regions consist of codes already described by HSCs. 

While we haven't ruled out the possibility of some highly irregular graph combining 3 and 4-valent vertices providing useful 2D stabilizer codes, we have shown that no semi-regular graphs will suffice.

\end{appendix}

\end{document}